\documentclass[notitlepage]{revtex4-1}

\usepackage{amsmath}
\usepackage{amssymb}
\usepackage{graphicx}
\usepackage{dcolumn}
\usepackage{bm}
\newcommand{\squeezedown}{\vspace{17mm}}

\usepackage[skip=-40pt]{caption}
\usepackage{epsfig,psfrag,graphicx,bm,color}
\usepackage{graphicx}
\usepackage{mathptmx}      
\usepackage{latexsym}
\def\beq{\begin{equation}}
\def\eeq{\end{equation}}
\def\bea{\begin{eqnarray}}

\def\eea{\end{eqnarray}}

\begin{document}

\title{Holographic entanglement entropy for charged accelerating AdS black holes}

       
\author{Masoumeh Tavakoli$^{1,2,3}$} \thanks{email: m.tavakoly@ph.iut.ac.ir} \,
  \author{ Behrouz Mirza$^{1}$ } \thanks{email: b.mirza@cc.iut.ac.ir}\,  
  \author{Zeinab Sherkatghanad$^{1}$}  \thanks{email: sherkat.elham@gmail.com }


\affiliation{$^1$Department of Physics, Isfahan University of Technology, Isfahan 84156-83111, Iran\\
 $^{2}$Department of Physics and Astronomy, University of Waterloo, Waterloo, Ontario N2L 3G1, Canada. \\
 $^{3}$Research Institute For Astronomy and Astrophysics Of Maragha, Maragha, 55177-36698, Iran }

\begin{abstract}
We investigate the holographic entanglement entropy in the Rindler-AdS space-time to obtain an exact
solution for the corresponding minimal surface. Moreover, the holographic entanglement entropy of the
charged single accelerated AdS Black holes in four dimensions is investigated. We obtain the volume of
the codimension one-time slice in the bulk geometry enclosed by the minimal surface for both the RindlerAdS space-time and the charged accelerated AdS Black holes in the bulk. It is shown that the holographic
entanglement entropy and the volume enclosed by the minimal hyper-surface in both the Rindler spacetime and the charged single accelerated AdS Black holes (C-metric) in the bulk decrease with increasing
acceleration parameter. Behavior of the entanglement entropy, subregion size and value of the acceleration
parameter are investigated. It is shown that for $|A| < 0.2$ a larger subregion on the boundary is equivalent
to less information about the space-time. 
\end{abstract}
\maketitle

\section{Introduction}
\label{sec:intro}

The Anti-de-Sitter/Conformal field theory correspondence is important for describing the gauge duality in which a strong conformal field theory (CFT) living in the boundary is dual to a weakly coupled gravity theory in Anti-de-Sitter (AdS) space-time in one higher dimension\cite{Maldacena, Gubser, Witten}. Ryu and Takayanagi  \cite{RT,RT1,BasCha} were the first to establish a relationship between entanglement entropy on the $d$ dimensional field theory side and the area of $d-1$ dimensional minimal surfaces in the bulk  anchored at the boundary of AdS space. There are usually UV divergences in the holographic entanglement entropy (HEE) whose removal requires a  method of regularization. Therefore, the holographic entanglement entropy for a deformed geometry can be subtracted from the term due to the background AdS space-time.

In Ref. \cite{HRT}, the covariant version of the
Ryu-Takayanagi formula for the time-dependent background was presented with the bulk metric
being neither static nor even stationary. Recently, the thermodynamics of accelerating AdS black
holes as described by the so-called C-metric, and their generalizations have attracted a lot of
attention \cite{Kubiznak1, Kubiznak2, Dias, Griffiths, Astron, Hong, HaoXu, Plebanski}. Ref. \cite{Plebanski} 
introduced a C-metric with mass and electric charge. The C-metric
were obtained by Walker and Kinnersley \cite{kinner}, and Bonnor \cite{bonnor}is a special space-time, which
include boost and rotation symmetries. In this space, the black holes contain both the black hole
horizon and the accelerating horizon. These black holes pulled with a cosmic strut or pushed by a
cosmic string. Also the acceleration is indicated with conical singularities in this space-time that
connect the event horizon of the black hole to infinity. When we set $m=0$ the spacetime is exactly
pure AdS. Then there is a coordinate transformation that brings it to the standard Rindler-AdS
form. One can then see that the parameter $A$ is the acceleration of the Rindler observer, for more details see \cite{Podolsky}.  
 For accelerating black holes, a negative cosmological constant can remove the
accelerating horizon as the second horizon to yield what has come to be called the "slowly accelerating black holes" \cite{Podolsky,Podolsky2}, and therefore the observer does not observe the Unruh temperature.
A clear visual representation of the global structure of the space-times for accelerated black holes
in the AdS space-time is investigated in \cite{Krtous}.
Also, it has been shown that a certain solution exists which represents  a pair of accelerated black holes if  the inverse of the cosmological length $\ell$ becomes smaller than  the acceleration parameter $A$
\cite{Dias}. In this condition, the accelerating observer observes the Unruh temperature \cite{Unruh}. At $A=1/\ell$ the accelerating horizon is extremal and the Unruh temperature is equal to zero (T=0). 

The thermodynamical properties of such an accelerating charged AdS black hole, i.e., the $P-V$ criticality and phase transition, are investigated in Ref. \cite{PVacc}.
The thermodynamics of accelerating black holes
 is also  studied in \cite{Kubiznak1}.
 Finally, the first law of thermodynamics for accelerating black holes has been recently obtained  \cite{Kubiznak2}.

In this paper, we obtain an exact solution for the minimal surface  in the Rindler-AdS space-time.
 We compute the holographic entanglement entropy associated with the minimal surface for the Rindler or the accelerated observer.
 It is worth figuring out how the holographic entanglement entropy behaves as a function of the acceleration parameter.
 Also, it will be interesting to determine numerically the holographic entanglement entropy of four-dimensional charged single accelerated AdS Black holes.
  We also  calculate the volume enclosed by the minimal surface.
We observe that the holographic entanglement entropy and the volume enclosed by the minimal hyper-surface decrease as a result of increasing acceleration parameters in both the Rindler-AdS space-time and the charged single accelerated AdS Black holes in the bulk \cite{Myers,fidelity,Alishahiha,Alishahiha1,LM,Susskind,Susskind2,Susskind3,Susskind4,CA1,CA2,CA3,CA4}.
 Finally, it must be noted that entanglement entropy plays a central role in quantum information theory \cite{QI3,QI1,QI2,QI4}.

The rest of the paper is organized as follows.
In Section \ref{part2}, an exact solution is obtained for the equation of motion related to  the holographic entanglement entropy for the Rindler-AdS space-time.
 Also, the volume enclosed by the minimal hyper-surface of the Rindler-AdS space-time is evaluated as a function of the acceleration parameter.
 In Section \ref{part3}, we review the thermodynamics of charged single accelerating AdS black holes.
 Moreover, we obtain both the holographic entanglement entropy and  the volume enclosed by the minimal hyper-surface, or the  Ryu-Takayanagi-volume (RT-volume),  for charged single accelerating AdS Black holes in the bulk.
\section{Exact solution related to HEE in the Rindler-AdS space-time}\label{part2}

To calculate the holographic entanglement entropy (HEE), we assume a time slice located at the AdS space-time and derive the minimal surface called $\gamma_A$. In this condition, the entanglement entropy is given by

\begin{equation}
\label{En}
S_A=\frac{Area(\gamma _A)}{4 G_{d+1}},
\end{equation}
where, G is the Newton's constant \cite{RT}. In order to evaluate  the minimal surface area in the bulk, we solve the Euler-Lagrange equation due to  the surface area \cite{HaoXu}. The holographic entanglement entropy for the Rindler-AdS(RAdS) space-time
is computed using the metric represented in Eq. (\ref{Rmetric}).
\begin{equation}
\label{Rmetric}
ds^2 =\frac{1}{\Omega ^2} \left[- f(r) dt^{2}+\frac{dr^2}{f(r)} +r^2(d\theta ^2+ \sin^2 \theta d\phi^2)\right] ,
\end{equation}
where,
\begin{eqnarray}
f(r) = 1-A^2 r^2+\frac{r^2}{\ell ^2} \,,
\\
\Omega =1+A \ r \cos \theta  \,.
\end{eqnarray}
Although this space-time is a locally pure AdS and there is no conical singularity, its boundary is located at $r=-1/(A \cos \theta)$ rather than at infinity.
In what follows, we compute the HEE  by using a new exact solution for the minimal surface
 in the Rindler-AdS  space-time.
 \\

 \begin{eqnarray}
Area =\int _{\theta =0}^{\theta _0}\int _{\phi =0}^{2\pi} \sqrt{det h}\ d\theta d\phi ,
\end{eqnarray}
  where the minimal surface is a time slice of AdS that is anchored on the entangling surface in the boundary of AdS.
 In order to evaluate  the minimal surface area in the bulk, we define an induced metric of the Rindler-AdS space-time on the surface as follows:
\begin{eqnarray}
&&h_{ab}=G_{\mu \nu}\frac{\partial x^\mu}{\partial \xi ^a}\frac{\partial x^\nu}{\partial \xi ^b}\,,~~~~~~~~~~~~~~~~~~~~~~~~~~~~
\nonumber\\
&&h_{11}=G_{rr}(\partial _{\theta}r(\theta))^2+G_{\theta \theta}=\frac{(\partial _{\theta}r(\theta))^2}{\Omega ^2 f(r(\theta))}+\frac{ r(\theta)^2}{\Omega ^2}\,,~~~~~~~~~~~~~~~
\nonumber\\
&&h_{22}=G_{\phi \phi}=\frac{ r(\theta)^2 \sin^2 \theta}{\Omega ^2}\,,~~~~~~~~~~~~~~~~~~~~~
\nonumber\\
&&h_{12}=h_{21}=0 \,,~~~~~~~~~~~~~~~~~~~~~~~~~~~~~~~~~
\nonumber\\
&&ds_{\gamma_A}^2 = \frac{1}{\Omega ^2} \left[\Big(\frac{\partial _{\theta} r(\theta)^2}{f(r(\theta))} +r^2(\theta)\Big)d\theta ^2+r^2(\theta) \ \sin ^2 \theta d\phi ^2\right] \,,
\end{eqnarray}
where,
 $r$  is a function of $\theta$.  Therefore, the area  from the above induced metric is
\begin{eqnarray}
\label{area}
Area =2 \pi \int_{0}^{\theta_0} \frac{r(\theta) \sin \theta }{\Omega ^2} \sqrt{\frac{r'^2(\theta)}{f(r(\theta))} + r^2(\theta)} \,,
\end{eqnarray}
where,
\begin{eqnarray}
&&f(r(\theta))=1-A^2 r(\theta)^2+\frac{r(\theta)^2}{\ell ^2}\,,
\\
&&\Omega =1+A \ r(\theta) \cos \theta \,.
\end{eqnarray}

We also have the following two boundary conditions in the Rindler-AdS space: $r'(\theta)=0$ and $r(\theta)=r_0$ at $\theta=0$ and  $r(\theta_0)=-1/(A \cos \theta_0)$. We regularize  the area by integrating out not to $\theta=\theta_0$ but to $\theta=\theta_0-\epsilon$, where $\epsilon$ is a UV cut-off. Consider the parameter $\theta$ as time in the following Lagrangian form
\begin{equation}
{\cal{L}} = \frac{r(\theta) \sin \theta}{\Omega ^2} \sqrt{\frac{r'^2(\theta)}{f(r(\theta))} + r^2(\theta)}\,.
\end{equation}
in which the prime indicates the derivative with respect to $\theta$. The Euler-Lagrange  equation  is given by:
\begin{equation}
\frac{\partial {\cal{L}} }{\partial r(\theta )}-\frac{\partial }{\partial \theta }(\frac{\partial {\cal{L}} }{\partial r'(\theta )})=0\,,
\end{equation}
this may be rewritten in the following form:
\begin{eqnarray}
\label{EOMn}
&&2 f(r(\theta )) r'(\theta )  \Big[ \Omega \sin \theta  r(\theta )^2 f'(r(\theta ))+r'(\theta )^2 (2 \sin \theta  \Omega '-\Omega  \cos \theta )\Big]
\nonumber\\
&&+\Omega  \sin \theta  f'(r(\theta) ) r'(\theta )^3-2 f(r(\theta) )^2 r(\theta ) \Big[ r(\theta ) \Big(\Omega  \sin \theta  r''(\theta )+
\nonumber\\
&&r'(\theta ) \left[\Omega  \cos \theta -2 \sin \theta  \Omega ' \right] \Big)-3 \Omega 
 \sin \theta  r'(\theta )^2 \Big]+4 f(r(\theta) )^3 \Omega  \sin \theta  r(\theta )^3
 \nonumber\\
 &&=0\,.
\end{eqnarray}
 which represents the exact solution of the Equation of motion (EOM) in the pure AdS space-time.
It is worthwhile obtaining an exact solution of the Euler-Lagrange equation for the Rindler-AdS space yielding the entanglement entropy associated with the minimal surface on the boundary in the  pure Rindler-AdS space-time. As explained in detail in Appendix, we obtain an exact solution for the above equation of motion by using the coordinate transformations as follows:
\begin{eqnarray}
\label{rthAdS}
&&r_{RAdS}(\theta)=\Big[2 A \ell^2 \cos \theta-2 A^3 \ell^4 \cos \theta +
 \nonumber\\
&&\sqrt{\frac{2 \ell^2 \left(A^2\ell^2-1\right) \cos ^2 \theta _0 \left(\frac{4 \cos ^2 \theta _0}{A^2 \cos \left(2 \theta _0\right)-A^2+2}+\left(A^2\ell^2-1\right) \cos (2 \theta )-A^2 \ell^2-1\right)}{A^2 \cos ^2 \theta _0-A^2+1}} \Big]/
 \nonumber\\
&&\Big[2 \left(A^2 \ell^2-1\right) \left(\frac{2 \cos ^2 \theta _0}{A^2 \cos \left(2 \theta _0\right)-A^2+2}+\left(A^2 \ell^2-1\right) \cos ^2 \theta _0\right)\textbf{\Big]}.
\end{eqnarray}
If we expand $r(\theta)$ about the acceleration parameter, A, we have
\begin{eqnarray}
&&r_{RAdS}(\theta)=\frac{\ell}{\sqrt{\frac{\cos ^2 \theta }{\cos ^2 \theta _0}-1}}+\frac{\ell^2 \cos \theta }{\cos ^2 \theta-\cos ^2 \theta _0} A+\frac{ \ell \cos \theta _0}{4 (\cos ^2 \theta -\cos ^2 \theta _0 )^{3/2}}\times
\nonumber\\
&&\Big[(4 \ell^2+1) (2 \cos ^2\theta -1)-(2 \cos ^2\theta _0-1) (2 \cos ^2 \theta +2 \ell^2)+6 \ell^2+1\Big] A^2
\nonumber\\
&&+O(A^3)\,.
\end{eqnarray}
 \begin{figure}
\squeezedown
  \includegraphics[width=8cm,height=5cm]{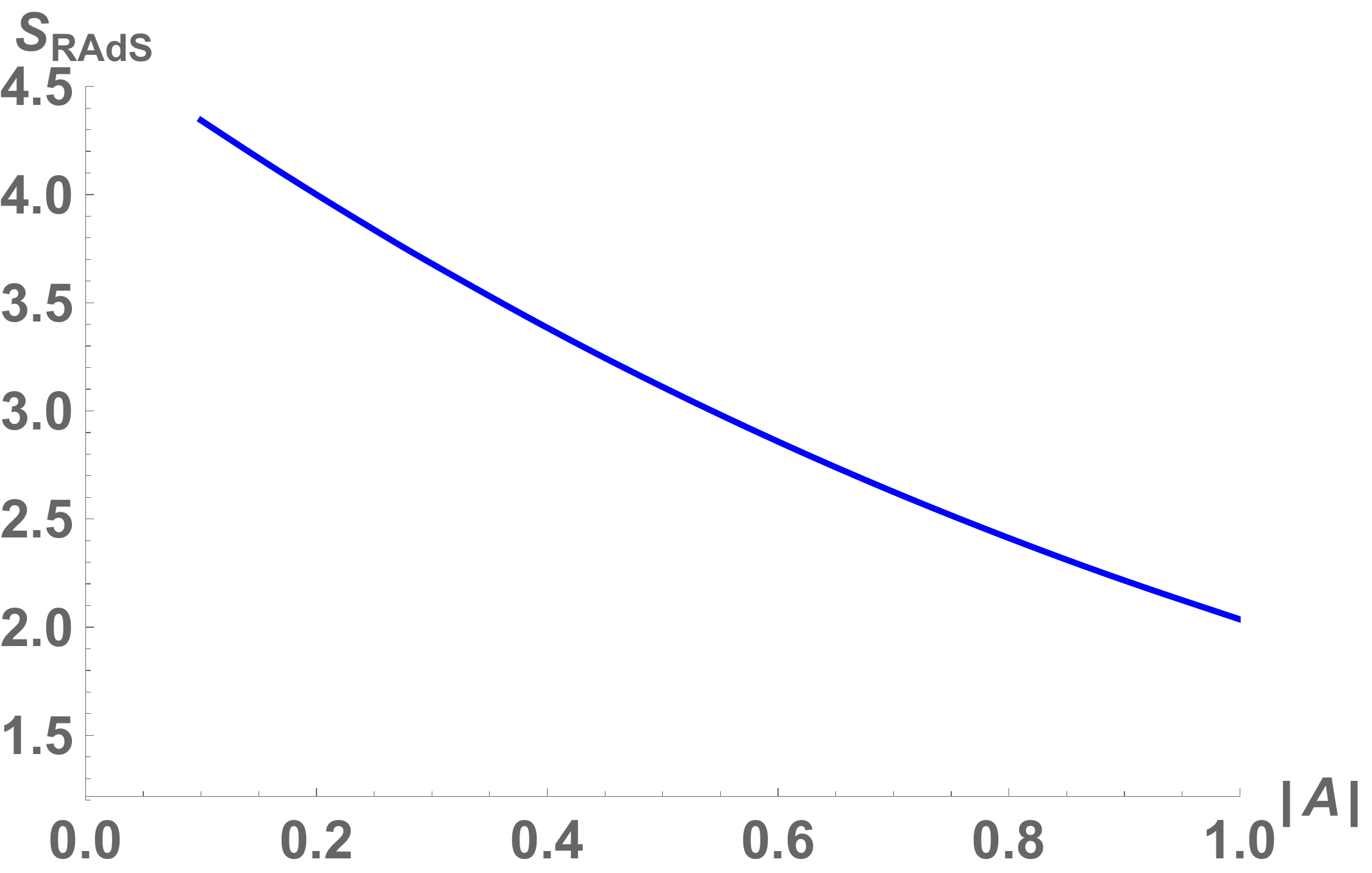}\\
\squeezedown
  \caption {\it{ The holographic entanglement entropy for the Rindler-AdS space, $S_{RAdS}$, with respect to the acceleration parameter, $A$, for $\theta_0=0.16$ and $\ell=1$. The holographic entanglement entropy decreases
with increasing  acceleration parameter. This result is  similar to the  behaviour of entanglement entropy in accelerating frames  studied in relativistic quantum information.}}\label{figure:fig1s}
 \end{figure}
The first term is exactly equal to the well-known result for the pure AdS space-time presented in \cite{Johnson}. The higher order terms indicate the effects of the acceleration parameter on the exact solution of $r(\theta)$ order by order. This new exact  solution can be used to obtain non-perturbatively the behavior of entanglement entropy as a function of acceleration parameter.\\
By substituting $r(\theta)$ from Eq. (\ref{rthAdS}) into Eq. (\ref{area}), we may obtain the minimal  surface area and, then, the holographic entanglement entropy from Eq.  (\ref{En}).  The holographic entanglement entropy is plotted with respect to the acceleration parameter $A$ in Fig. \ref{figure:fig1s}. Throughout this paper, we take $\ell=1$, $\theta_0=0.16$, and $\epsilon=0.005$. In Fig. \ref{figure:fig1s}, the holographic entanglement entropy in the Rindler-AdS space, $S_{ RAdS}$, decreases  with increasing  acceleration parameter, $A$, in the range of $A < 1/\ell$. Thus, a loss of information and the corresponding degradation of entanglement entropy occur with increasing  acceleration parameter, A.  This means that the holographic entanglement entropy follows the universal behavior of entanglement entropy  obtained in single mode studies related to accelerating observers in the relativistic quantum information perspective \cite{QI3,QI1,QI2,QI4}.\\
\begin{figure}[tbp]
\centering 
  \includegraphics[width=8cm,height=8cm]{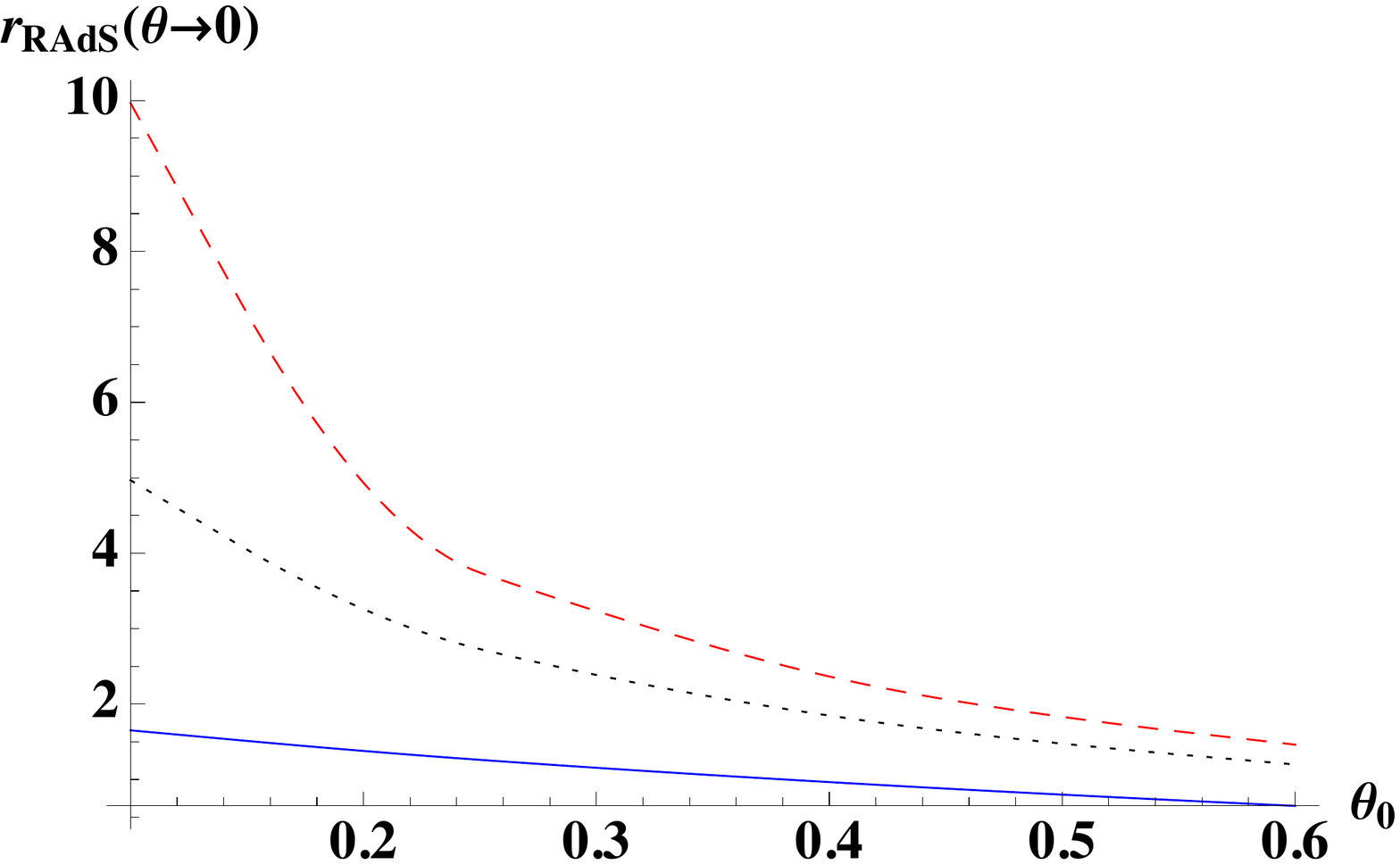}\\ 
\caption{\label{figure2} The solution $r_{RAdS}(\theta)$ near $\theta =0 $ with respect to  $\theta_0$, for $|A|=0$ (dashed \ (red) \ line), $0.1$ (dotted  \ (black) \ line), $0.5$ (solid  \ (blue) \ line) and $\ell=1$.}
\end{figure}
In Fig. \ref{figure2}, the relationship between $r_{RAdS}(\theta \rightarrow 0)$ and $\theta _0$ is described, where $r_{RAdS}(\theta \rightarrow 0)$ decreases when $\theta _0$ or the subregion in the boundary increased. Also we explore the entanglement entropy as function of $\theta _0$ for different values of the accelarating parameter A in (Fig.  \ref{figure3}). It should be noted that a threshhold value exists for  the accelarating parameter A equal to $|A|=0.2$. 
For $|A|<0.2$, by increasing entanglement entropy,  $\theta _0$ increases and for $|A|>0.2$, increasing entanglement entropy, $\theta _0$ decreases. Thus for $|A|<0.2$ a larger subrregion on the boundary is equivalent to less information about the space-time. However, this result is vice versa for $|A|>0.2$.
\noindent We consider a volume as the codimension one-time slice in the bulk geometry enclosed by the minimal area
 in the Rindler-AdS space, or  the so-called "Ryu-Takayanagi volume (RT-volume)", represented by:
\begin{eqnarray}
\label{VAdS}
V_{RAdS}=2 \pi \int_0 ^{\theta_0} \sin \theta \ d \theta \int_{r_{RAdS}(0)}^{r_{RAdS}(\theta)} \frac{r^2 \ d r}{\sqrt{1-A^2+\frac{r^2}{\ell ^2}} \ (1+A r \cos \theta )^3} \,,
\end{eqnarray}
by substituting  $r(\theta$) from the Euler-Lagrange equation in Eq. (\ref{rthAdS}), we might obtain the RT volume enclosed by the minimal  surface area. We assume the size of the boundary region to be  $\theta_0 = 0.16$ and $\epsilon=0.005$ and compute the above integral numerically.

The results show that the volume enclosed by the minimal hyper-surface decreases
as the acceleration parameter increases in the Rindler space-time. The result is depicted in Fig.  \ref{figure4}. It is interesting to find an equivalent quantity for RT volume in the relativistic Quantum Information Theory.
 \begin{figure}
\squeezedown
  \includegraphics[width=8cm,height=5cm]{{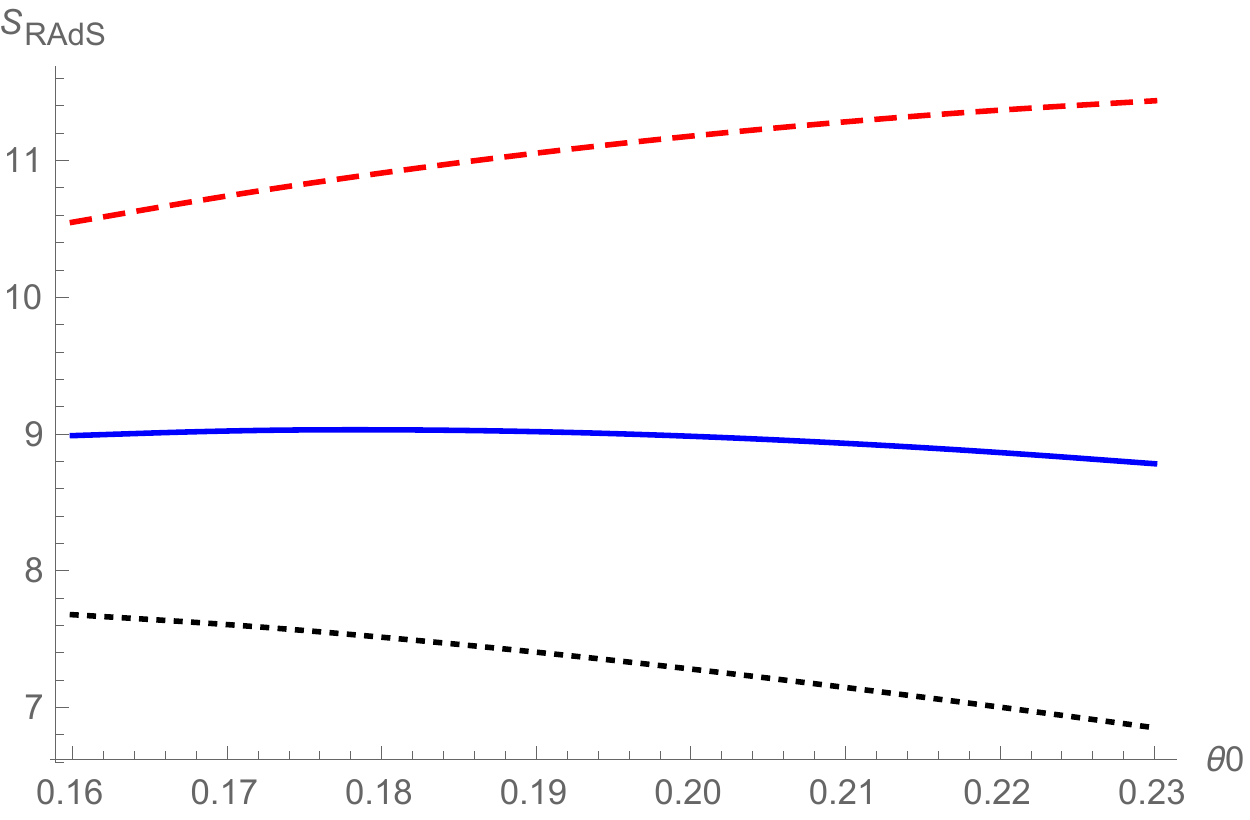}}\\ 
\squeezedown
  \caption {\it{The holographic entanglement entropy for the Rindler-AdS space, $S_{RAdS}$, with respect to $\theta_0$, for $|A|=0.1$ (dashed \ (red) \ line), $0.2$ (solid  \ (blue) \ line), $0.5$ (dotted  \ (black) \ line) and $\ell=1$. The holographic entanglement entropy decreases
with increasing  $\theta_0$ parameter.}} \label{figure3} 
\end{figure}
 \begin{figure}
\squeezedown
  \includegraphics[width=8cm,height=5cm]{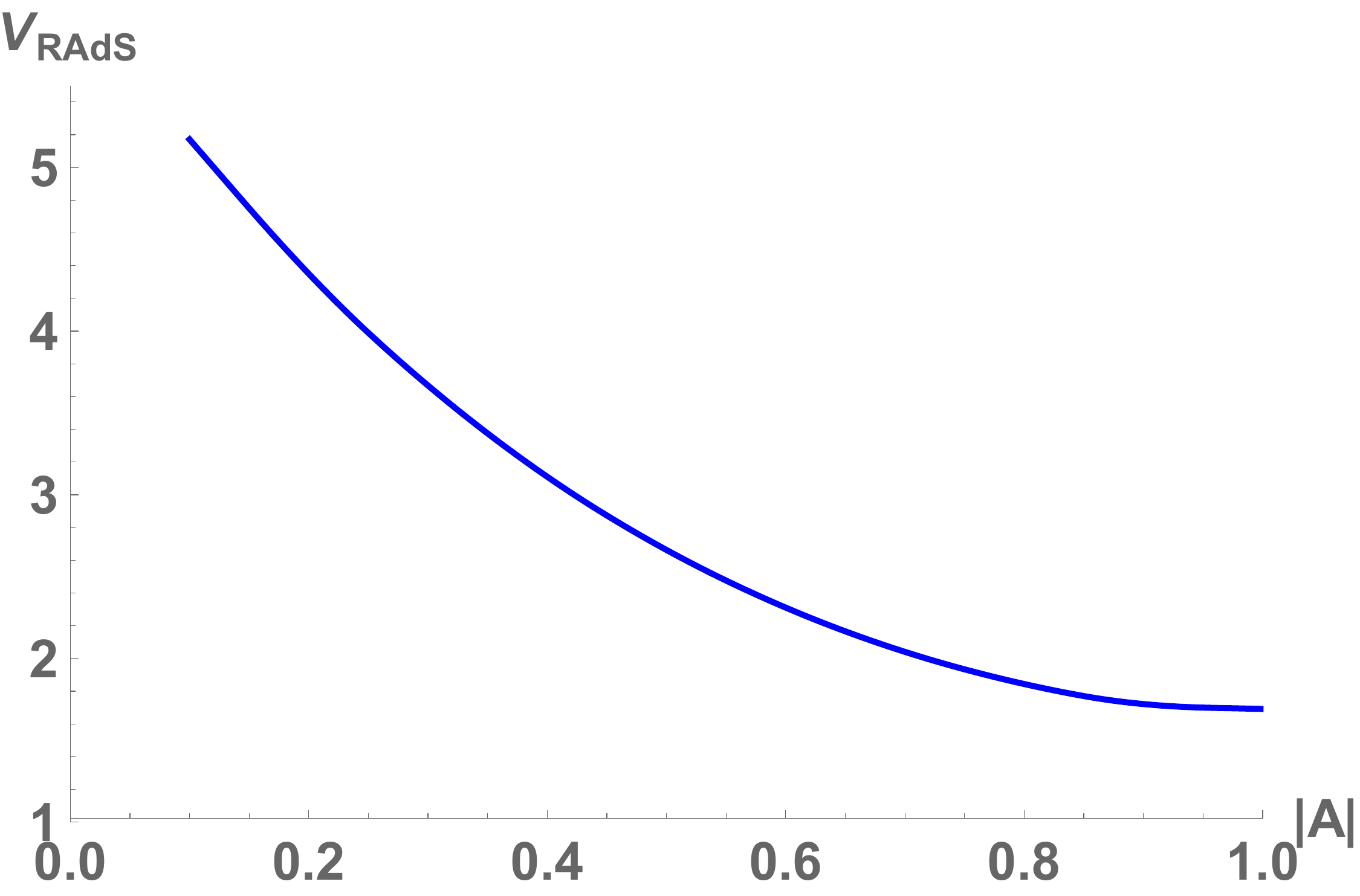}\\
\squeezedown
  \caption {\it{ RT-volume for the Rindler-AdS space-time, $V_{RAdS}$, with respect to the acceleration parameter $A$ for the Rindler-AdS space-time where $\theta_0=0.16$. 
  The volume decreases for  larger values of  acceleration.
  }}\label{figure4}
 \end{figure}
\section{Holographic entanglement entropy of charged accelerating black holes}\label{part3}
In this Section, we review the metric of slowly accelerating Black Holes in the anti-de Sitter space.
Ref. \cite{AFV} studied the black hole with a cosmic string in which a conical deficit is considered in a Schwarzschild black hole. The thermodynamics of this type of black hole  under varying tensions was studied in \cite{Kubiznak2}.
A more general geometry that corresponds to accelerated black holes
is described by C-metric.
The usual form of C-metric as reported in \cite{kinner,Plebanski} represents either one or two accelerating black holes.
 The accelerating and rotating black hole space-time was studied in \cite{Podolsky2}
 while Ref. \cite{Chen} computed the motion of timelike particles along geodesics.
 
In the present work, we ignore the acceleration horizons of  C-metrics  and consider only the black hole horizon to  have a well-defined temperature.
 \begin{figure}
\squeezedown
  \includegraphics[width=8cm,height=5cm]{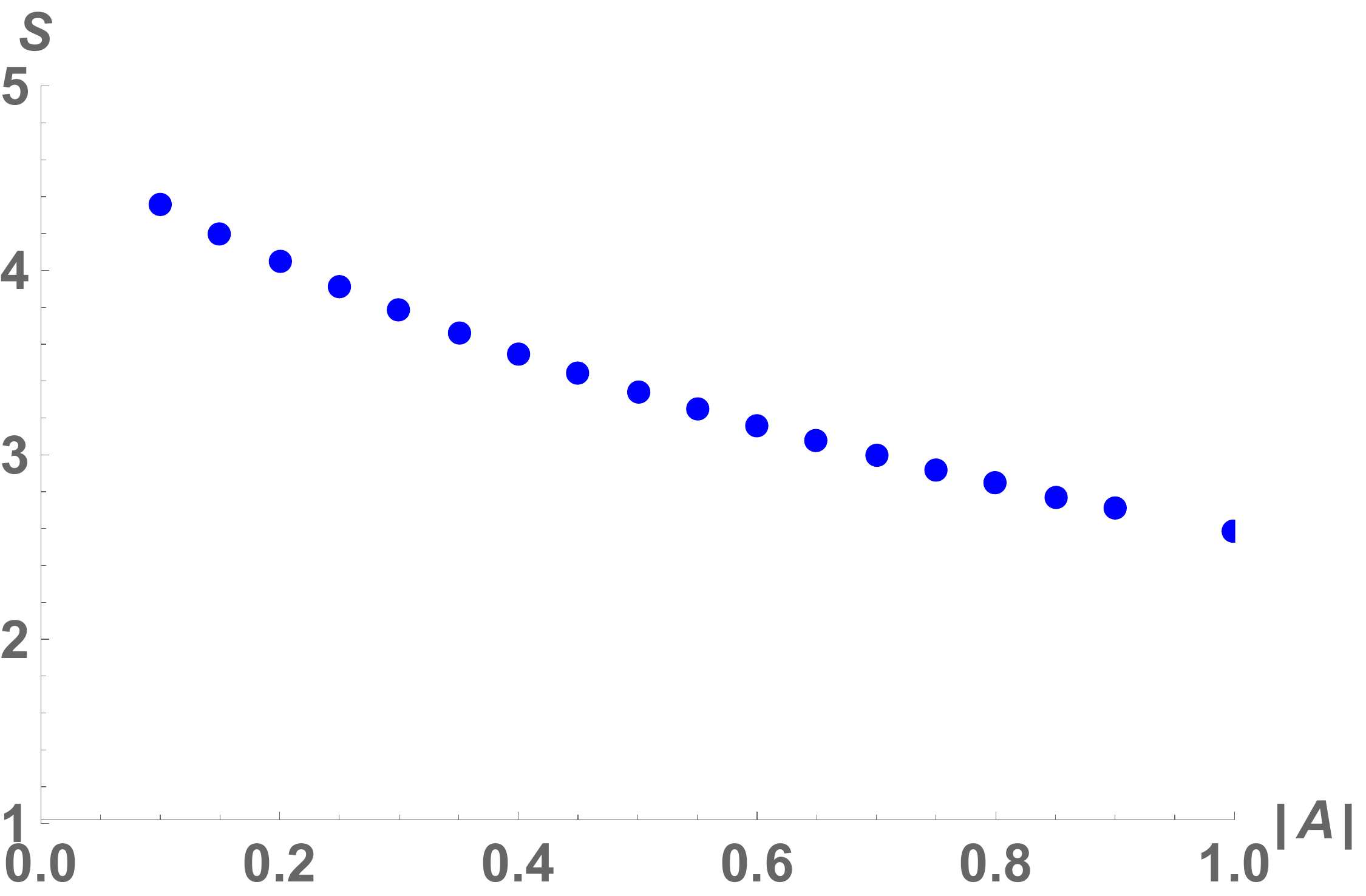}\\ 
\squeezedown
  \caption {\it{ The holographic entanglement entropy for the charged Accelerating AdS Black hole with respect to the acceleration parameter $A$. Here, we take $e=0.01$, $m=0.1$, $\ell=1$ and $\theta_0=0.16$. The holographic entanglement entropy
   decreases
with increasing  acceleration to exhibit  behaviour similar to that observed in the earlier studies of relativistic quantum information.}}
\label{figure5}
\end{figure}
 This simplification is known as the "slowly accelerating C-metric" \cite{Podolsky2}.
 One can show that, for $A < 1/\ell$, a single black hole  appears in the AdS with the only horizon being that of the black hole. For $A > 1/\ell$, we have two black holes separated
by the acceleration horizon \cite{Dias,Podolsky2,Krtous}. Typically, a C-metric with a cosmological constant and charge in the Hong-Teo coordinates is represented by the following metric \cite{Griffiths,Hong}:
\begin{equation}
    \label{metric}
ds^2 =\frac{1}{\Omega ^2} \left[- f(r) \frac{dt^{2}}{\alpha ^2}+\frac{dr^2}{f(r)} +r^2 \big(\frac{d\theta ^2}{g(\theta)}+g(\theta) \sin^2 \theta \frac{d\phi ^2}{K^2}\big)\right],
\end{equation}
where,
\begin{eqnarray}
&&f(r)=(1-A^2 r^2)(1-\frac{2m}{r}+\frac{e^2}{r^2})+\frac{r^2}{\ell ^2}\,,
\\
&&\Omega =1+A \ r \cos \theta \,,
\\
&&g(\theta) =1+2 \ m \ A \cos \theta+e^2 A^2 \cos ^2 \theta \,,
\\
&&\alpha =\sqrt{ (1+A^2 e^2)(1-A^2 \ell^2 (1+A^2 e^2))}\,,
\end{eqnarray}
where, the two parameters  $m$ and $e$ represent the black hole's mass and charge, respectively. $A$ is the magnitude of acceleration of the black hole and $l$ is the cosmological length. Also,  $\Omega$ as the conformal factor modifies the location of the AdS boundary from infinity $r=\infty$ to finite values $r=\frac{-1}{A \cos \theta}$. This space-time has conical singularities  $(\theta=0,\pi)$ and the regularity condition of
the metric at the poles, $\theta_+=0$ and $\theta_-=\pi$ require that:
\begin{equation}
  K_{\pm}=g(\theta_{\pm})=1+2 m A+e^2 A^2.
\end{equation}
In general, $K$ is rendered to regularize one pole, which leads to a conical deficit or a conical excess along the other pole. Since a negative energy object is the source of a conical excess, it is assumed throughout this paper that the black hole is regular on the North Pole where $\theta=0$ and $K =K_+$. Thus,  a conical deficit exists on the other pole where $\theta=\pi$. The first law of thermodynamics for accelerating black holes with a varying conical deficit and a critical behavior  is studied in \cite{Kubiznak1}.

The entropy of the black hole and the relevant temperature 
are given by
\begin{eqnarray}
&&S=\frac{\pi r_+^2}{K (1-A^2 r_+^2)}\,,
\\
&&T=\frac{f'(r_+)}{4 \pi \alpha}=\frac{1}{2 \pi r_+^2 \alpha} \Big[(1-A^2 r_+^2)(m-\frac{e^2}{r_+})+\frac{r_+^3}{l^2 (1-A^2 r_+^2)}\Big] \,.
\end{eqnarray}
 \begin{figure}
\squeezedown
  \includegraphics[width=8cm,height=5cm]{{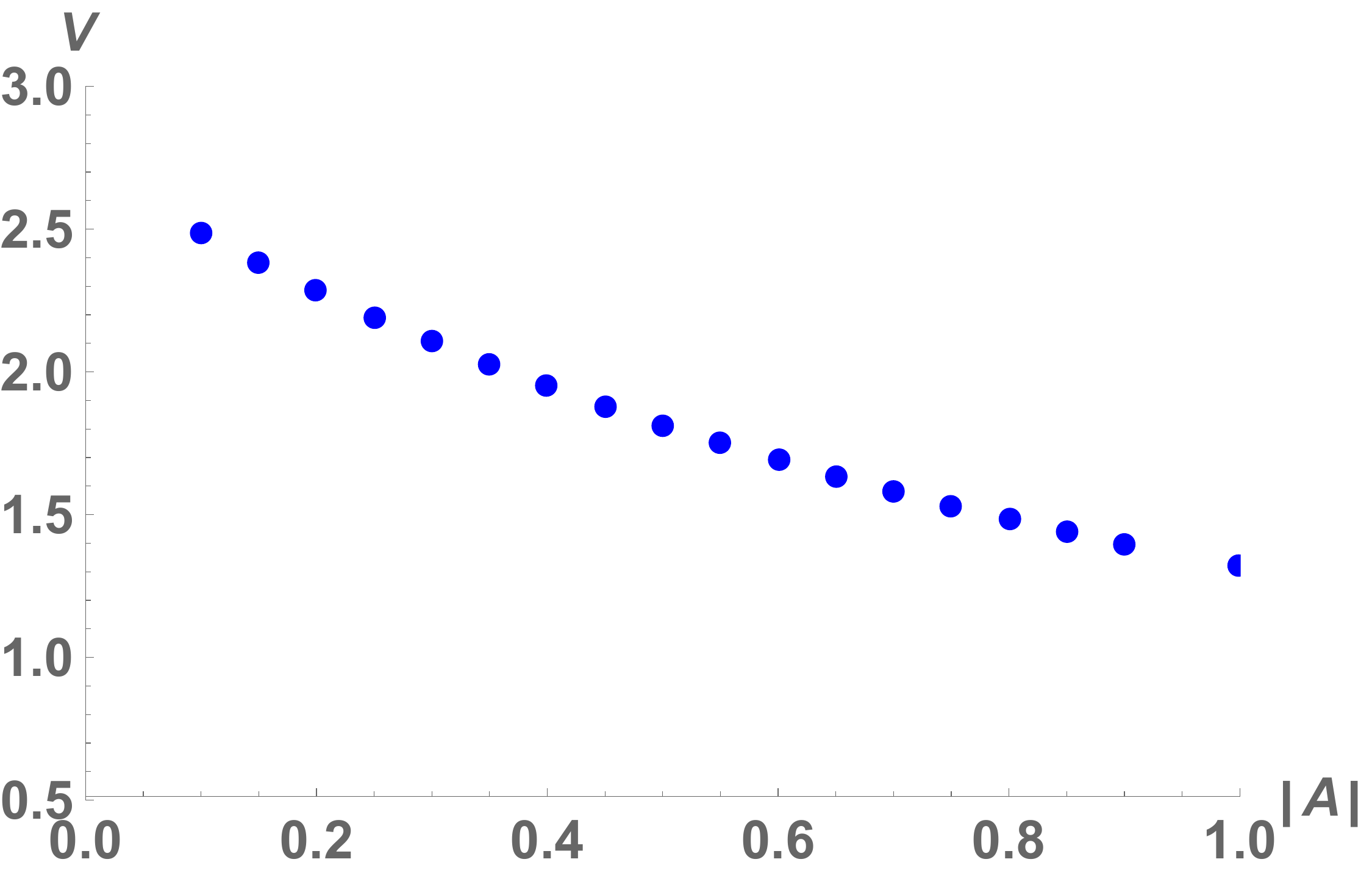}}\\ 
\squeezedown
  \caption {\it{ 
The RT-volume or subregion complexity, $V$, for a charged accelerating AdS Black hole with respect to the acceleration parameter $A$. Here, we take $e=0.01$, $m=0.1$, $\ell=1$ and $\theta_0=0.16$. The volume  decreases as a function of the acceleration parameter.}}
\label{figure6} 
\end{figure}
\noindent The holographic entanglement entropy for charged accelerated black holes can be evaluated using Eq. (\ref{En}) and the metric presented in Eq. (\ref{metric}). The area associated with the  minimal hyper-surface in the presence of a charged accelerating AdS Black hole in the bulk is captured by:
\begin{equation}\label{A}
Area =2 \pi \int_{0}^{\theta_0} \frac{r(\theta) \sin\theta \ \sqrt{g(\theta)}}{\Omega ^2 K} \sqrt{\frac{r'(\theta)^2}{f(r(\theta))} + \frac{r(\theta)^2}{g(\theta)}}\,,
\end{equation}
where,
\begin{eqnarray}
&&f(r(\theta))=(1-A^2 r(\theta)^2)(1-\frac{2m}{r(\theta)}+\frac{e^2}{r(\theta)^2})+\frac{r(\theta)^2}{\ell ^2}\,,
\\
&&\Omega =1+A \ r(\theta) \cos \theta\,,
\\
&&g(\theta) =1+2 \ m \ A \cos \theta+e^2 A^2 \cos ^2 \theta\,,
\\
&&K =1+2 m A+e^2 A^2\,,
\end{eqnarray}
where, the parameters $e$ and $m$ are the black hole's   electric charge and mass, respectively. To compute the holographic entanglement entropy, we need to minimize this surface  area. Therefore, a Lagrangian associated with this  surface  area  to obtain the EOM  takes the following form:
\begin{equation}
{\cal{L}}= \frac{r(\theta) \sin\theta \ \sqrt{g(\theta)}}{\Omega ^2 K} \sqrt{\frac{r'(\theta)^2}{f(r(\theta))} + \frac{r(\theta)^2}{g(\theta)}}\,.
\end{equation}
The equation of  motion is then given by
\begin{eqnarray}
\label{rth2}
&&g(\theta )^2 \Omega  \sin \theta  f'(r(\theta) ) r'(\theta )^3+2 f(r(\theta) )^2 r(\theta ) \Big[r'(\theta ) \Big(2 g(\theta ) 
\nonumber\\
&&\sin \theta  r(\theta ) \Omega '-\Omega  \big[\sin \theta  r(\theta ) g'(\theta )+g(\theta ) (\cos \theta  r(\theta )-3 \sin \theta  
\nonumber\\
&&r'(\theta ) )\big]\Big)-g(\theta ) \Omega  \sin \theta  r(\theta ) r''(\theta )\Big]+4 f(r(\theta) )^3 \Omega  \sin \theta  r(\theta )^3
\nonumber\\
&& f(r(\theta) ) g(\theta ) r'(\theta ) \Big(2 \Omega  \sin \theta  r(\theta )^2 f'(r(\theta ))-r'(\theta )^2 [\Omega  \sin \theta 
\nonumber\\ 
&& g'(\theta )-4 g(\theta ) \sin \theta  \Omega '+2 g(\theta ) \Omega  \cos \theta ]\Big)=0\,.
\end{eqnarray}
We may numerically solve the above equation of motion  to obtain $r(\theta)$. Here, the boundary conditions are $r'(\theta)=0$ and $r=r_0$ at $\theta=0$ and $r=-1/(A \cos \theta_0)$ at $\theta=\theta_0$. Substituting $r(\theta)$ from  Eq. (\ref{rth2}) in Eq. (\ref{A}), we can depict the holographic entanglement entropy with respect to the acceleration parameter A as shown in Fig.  \ref{figure5},
where $e=0.01$, $m=0.1$ and $\ell=1$.
Also, the Ryu-Takayanagi volume for charged accelerating AdS Black holes in the bulk is expressed by:
\begin{eqnarray}
\label{V}
&&V=2 \pi \int_0 ^{\theta_0} \ d \theta \int_{r(0)}^{r(\theta)} \ d r \Big[
\nonumber\\ 
&& \frac{r^2  \sin \theta}{(1+2 m A+e^2 A^2) \ (1+A \ r \cos \theta)^3 \sqrt{(1-A^2 r^2)(1-\frac{2m}{r}+\frac{e^2}{r^2})+\frac{r^2}{\ell ^2}}}\Big].
\end{eqnarray}
Replacing $r(\theta)$ from the Euler-Lagrange Eq. (\ref{rth2}) in Eq. (\ref{V}), we might plot the volume of the minimal hyper-surface with respect to $A$ as in Fig.  \ref{figure6}.
Clearly, the values of both HEE and the RT-volume of the minimal hyper-surface decrease as a function of the acceleration parameter when the charged accelerating AdS black hole is located in the bulk. This behavior is similar to those reported in studies of the entanglement entropy within the framework of the  relativistic quantum information theory.

\section{Conclusion}
 The entanglement entropy can be calculated for accelerating observers.
The usual form of a C-metric  represents either one or two accelerating black holes. In this study, we considered a C-metric with cosmological constant in the Hong-Teo coordinates to obtain  an exact solution for the minimal surface area. The holographic entanglement entropies of the Rindler-AdS space-time and charged single accelerated AdS Black holes were evaluated in four dimensions.
Moreover, the volume enclosed by the minimal area,
or the RT-volume, was computed for both the Rindler-AdS space-time and the charged single accelerated AdS Black holes. For this purpose, we considered an acceleration parameter less than the inverse of the cosmological length $\ell$, where a single black hole  appears in AdS with the only horizon being that of the black hole. Our exact solution of the Euler-Lagrange equation in the Rindler-AdS space-time is a novel result.

 Due to the appearance of a communication horizon, there is no possibility for a uniformly accelerated observer to access information about the whole of the space-time; hence,  a loss of information  occurs leading to  entanglement degradation. This implies that in a space-time with horizon one might expect some kind of  information loss puzzle. Efforts were then made to figure out how the holographic entanglement entropy behaves as function of the acceleration parameter.
  The holographic entanglement entropy and the the volume enclosed by the minimal hyper-surface were observed to decrease  with increasing acceleration parameters in the Rindler-AdS space-time, indicating that the holographic entanglement entropy follows the universal behaviour of entanglement entropy for accelerating observers.  It seems that existence of horizon in a space-time, entanglement entropy and information loss are forming a triangle with nontrivial relations.  

In addition, we extended our calculation to the charged single accelerating AdS Black holes in the bulk. The holographic entanglement entropy and volume (as the codimension one-time slice in the bulk geometry enclosed by the minimal  surface area)
  were computed numerically for charged accelerating AdS Black holes in the bulk to conclude that the holographic entanglement entropy and the volume enclosed by the minimal hyper-surface decrease   as a function of acceleration parameters. 
  
For future research, it  will be interesting to expand the calculation for  $A > 1/\ell$ with two black holes separated by the acceleration horizon.

\section{Acknowledgements}

This work has been supported financially by Research Institute for Astronomy and Astro
physics of Maragha (RIAAM) under research project No. 1/5750-52. We would like to thank
Robert B. Mann for comments and useful conversations.

\section{Appendix}

The general form of the metric that splits the $S^{d-1}$ into a polar angle $\Theta$ and a unit-radius sphere
$S^{d-2}$ with $d\Omega^2_{d-2}$ in $AdS_{d+1}$ is given by

\begin{equation}\label{metricapp}
ds^2=\frac{\ell^2}{\cos^2 \chi}\Big [-\frac{d T ^2}{\ell^2}+ d \chi^2+ \sin^2 \chi \big(d \Theta ^2+\sin^2 \Theta \ d\Omega^2_{d-2} \big) \Big]\,.
\end{equation}

In order to evaluate the entanglement entropy, a minimal hyper-surface ending on the boundary slice should be  considered. In this condition, we choose the following induced metric in which $\chi=\chi(\Theta)$ on the hyper-surface:

\begin{equation}
ds^2=\frac{\ell^2}{\cos ^2 \chi}\Big[\big((\partial_{\Theta} \chi)^2+\sin ^2\chi \big) d \Theta^2+\sin^2 \chi \sin^2 \Theta d\Omega^2_{d-2}\Big]\,.
\end{equation}

The entangling region on the boundary is a cap-like one, $0<\Theta<\Theta _0$. The surface area for the above induced metric is, therefore, expressed by

\begin{eqnarray}
Area=l^{d-1} Vol(S^{d-2}) \int d \Theta \frac{(\sin \chi \sin \Theta)^{d-2}}{\cos ^{d-1} \chi} 
 \sqrt{(\frac{d \chi}{d \Theta})^2+\sin^2 \chi}\,.
\end{eqnarray}

In order to minimize this  surface area, which is stable for small deformations, one needs to solve the resulting Euler-Lagrange equation. The boundary conditions are given by:
\begin{equation}
\cos \Theta_0=b   \   \   \   \    at   \   \    \chi=\pi/2 \,.
\end{equation}
The solution of this  equation of motion is not simple but it has been shown  in Ref. \cite{app} that the following function is satisfied in the Euler-Lagrange equation
\begin{equation}\label{sol}
\chi(\Theta)=\sin ^{-1} (\frac{b}{\cos(\Theta)}).
\end{equation}
If  a coordinate transformation is inserted into Eq. (\ref{metricapp}):
\begin{eqnarray}\label{CT}
&&\cos  \chi  = \frac{\sqrt{1-\ell^2 A^2}+A R \cos \xi}{\sqrt{1+R^2/\ell^2}},
\nonumber\\
&&l \tan \chi \sin \Theta=\frac{R \sin \xi}{\sqrt{1-\ell^2 A^2}+A \ R \cos \xi},
\nonumber\\
&&l \tan \chi \cos \Theta=\frac{\sqrt{1-\ell^2 A^2} R \cos \xi - l^2 A}{\sqrt{1-\ell^2 A^2}+A R \cos \xi},
\end{eqnarray}
we get the metric in  the Rindler-AdS space-time as follows:
\begin{eqnarray}\label{37}
&&ds^2 =\frac{1}{\Omega ^2} \left[- f(R) dT^{2}+\frac{dR^2}{f(R)} +R^2(d\xi ^2+\sin ^2 \xi d\phi^2)\right]\,,
\\
&&f(R)=1+\frac{R^2}{\ell^2}\,,
\\
&&\Omega =\sqrt{1-\ell^2 A^2}+A \ R \ \cos \xi \,.
\end{eqnarray}
Alternatively, inserting the following coordinate transformation  in Eqs. (\ref{37}):
\begin{equation}\label{CT1}
1+\frac{R^2}{l^2}=\frac{1+(1-A^2 l^2) \ r^2 \ell ^2}{(1-A^2 \ell^2)\Omega ^2 }  \   \   \  \  and   \   \   \   \  R\sin \xi= \frac{r \sin  \theta}{\Omega},
\end{equation}
will yield the  form of the metric  used  throughout the paper for the Rindler-AdS space-time:
\begin{eqnarray}\label{MM}
    &&ds^2 =\frac{1}{\Omega ^2} \left[- f(r) dT^{2}+\frac{dr^2}{f(r)} +r^2\big(d\Theta ^2+\sin ^2 \theta d\phi ^2\big)\right]\,,
    \\
   &&f(r)=1+\frac{r^2}{\ell ^2}(1-A^2 \ \ell ^2)\,,
   \\
   &&\Omega =1+A r \cos \theta \,.
\end{eqnarray}
Replacing  the coordinate transformations in Eqs. (\ref{CT}) and (\ref{CT1}) into the solution of the Euler-Lagrange equation in Eq. (\ref{sol}), we obtain the solution of the  Euler-Lagrange equation from the metric in Eq. (\ref{MM}) as follows:
\begin{eqnarray}
\label{rth}
&&r_{RAdS}(\theta)=\Big[2 A \ell^2 \cos \theta-2 A^3 \ell^4 \cos \theta +
\nonumber\\
&&\sqrt{\frac{2 \ell^2 \left(A^2 \ell^2-1\right) \cos ^2 \theta _0 \left(\frac{4 \cos ^2 \theta _0}{A^2 \cos \left(2 \theta _0\right)-A^2+2}+\left(A^2 \ell^2-1\right) \cos (2 \theta )-A^2 \ell^2-1\right)}{A^2 \cos ^2 \theta _0-A^2+1}} \Big]/
\nonumber\\
&&\Big[2 \left(A^2 \ell^2-1\right) \left(\frac{2 \cos ^2 \theta _0}{A^2 \cos \left(2 \theta _0\right)-A^2+2}+\left(A^2 \ell^2-1\right) \cos ^2 \theta _0\right)\Big]\,.
\end{eqnarray}
Thus, using the above methodology and a coordinate transformation in the metric in Eq. (\ref{MM}), an exact solution may be obtained for the complicated form of the Euler-Lagrange equation in Eq. (\ref{EOMn}) resulting from the induced  surface area. This provides the possibility to calculate not only the RT-entanglement entropy but also the RT-volume for the Rindler-AdS space-time non-perturbativly.
\section{Acknowledgements}

This work has been supported financially by Research Institute for Astronomy and Astro
physics of Maragha (RIAAM) under research project No. 1/5750-52. We would like to thank
Robert B. Mann for comments and useful conversations.


\end{document}